\pdfoutput=1

\documentclass[11pt]{article}

\usepackage[final]{acl}

\usepackage{times}
\usepackage{latexsym}

\usepackage[T1]{fontenc}

\usepackage[utf8]{inputenc}

\usepackage{inconsolata}

\usepackage{algorithm,algpseudocode}
\usepackage{microtype}
\usepackage{subfigure}
\usepackage{booktabs}
\usepackage{hyperref}
\usepackage{amsmath}
\usepackage{amssymb}
\usepackage{mathtools}
\usepackage{amsthm}
\usepackage{url}
\usepackage{colortbl}
\usepackage{listings}
\usepackage[dvipsnames]{xcolor}
\usepackage{wrapfig}
\usepackage{graphicx}
\usepackage{subcaption}
\usepackage{xspace}
\usepackage{multicol, multirow, bigstrut}
\usepackage{verbatim}
\usepackage{mdframed}
\usepackage{adjustbox}
\usepackage{tabularx}
\usepackage{enumitem}
\usepackage{bm}

\usepackage{tcolorbox}
\tcbuselibrary{listings}
\definecolor{shadecolor}{gray}{.9}
\definecolor{Gray}{gray}{0.9}
\newcommand{\our}[1]{\textsc{RethinkMCTS}}

\usepackage[capitalize,noabbrev]{cleveref}

\title{RethinkMCTS: Refining Erroneous Thoughts in Monte Carlo Tree Search for Code Generation}

\author{
 \textbf{Qingyao Li\textsuperscript{1}},
 \textbf{Wei Xia\textsuperscript{2}},
 \textbf{Xinyi Dai\textsuperscript{2}},
 \textbf{Kounianhua Du\textsuperscript{1}}, \\
 \textbf{Weiwen Liu\textsuperscript{1}},
  \textbf{Yasheng Wang\textsuperscript{2}},
 \textbf{Ruiming Tang\textsuperscript{2}},
 \textbf{Yong Yu\textsuperscript{1}},
\textbf{Weinan Zhang\textsuperscript{1}\thanks{Corresponding authors.}}, 
\\
 \textsuperscript{1}Shanghai Jiao Tong University,
 \textsuperscript{2}Huawei Noah's Ark Lab
\\
\texttt{\{ly890306,wnzhang\}@sjtu.edu.cn}  \\
}

\begin{document}
\maketitle
\begin{abstract}
Tree search methods have demonstrated impressive performance in code generation. Previous methods combine tree search with reflection that summarizes past mistakes to achieve iterative improvement. However, these methods face significant challenges. First, they search directly within the code language space, neglecting the underlying reasoning process critical for effective code generation. Second, reflection-based approaches merely accumulate historical errors in memory without providing correct reasoning pathways, making it difficult for subsequent search iterations to identify optimal solutions, resulting in decreased search quality.
In this work, we propose \our{}, a framework that systematically explores and refines the reasoning process for code generation. Specifically, we employ MCTS to search for thoughts before code generation and integrate MCTS with a refinement mechanism called \textit{rethink}, which incorporates fine-grained code execution feedback to refine erroneous thoughts during the search. It ensures the search path aligns with better reasoning, improving overall search quality. Through extensive experiments, we demonstrate that \our{} outperforms previous search-based and feedback-enhanced code generation baselines\footnote{Resources are available at \href{https://github.com/SIMONLQY/RethinkMCTS}{https://github.com/SIMONLQY/RethinkMCTS}.}.
\end{abstract}

\section{Introduction}
\label{intro}

With the impressive capabilities of large language models (LLMs), research has increasingly focused on enhancing their code generation abilities~\citep{luo2023wizardcoder, zheng2023self, gong2024ast}. 
Code generation is a reasoning task that requires multiple attempts and iterative corrections to achieve accurate results \citep{zhou2025refinecoder, bi2024iterative}, hence search algorithms demonstrate particular promise in this domain, achieving state-of-the-art performance \citep{delorenzo2024make, zhang2023planning, kulal2019spoc, zhou2023language}. 
Unlike other reasoning tasks, code environments provide rich execution feedback that can be leveraged to improve results. Previous approaches, such as LATS \citep{zhou2023language}, have effectively combined search with the reflection mechanism, enabling search trees to summarize past errors based on feedback and store them in memory to enhance subsequent search performance.
\begin{figure}[t]
    \centering
    \includegraphics[width=0.97\linewidth]{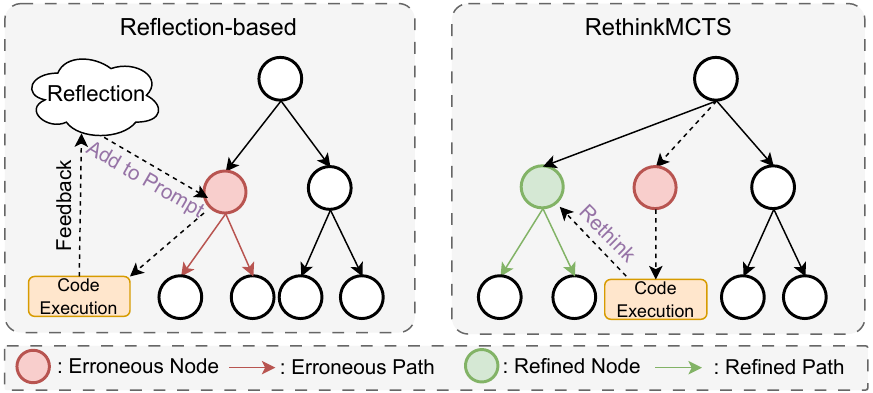}
    \caption{Comparison between reflection-based methods and \our{}. Reflection-based methods would maintain the error in the path, while \our{} would refine erroneous thoughts and continue along a better path.}
    \label{fig:intro}
\end{figure}

Despite demonstrating promising results, previous methods still face two key challenges: 1) Insufficient reasoning exploration. Studies, such as chain-of-thought~\citep{wei2022chain} and tree of thoughts~\citep{yao2024tree}, show that explicitly modeling the reasoning process leads to better results. ~\citet{tang2023large} further highlighted that LLMs are better equipped for semantic reasoning than symbolic reasoning. However, for code generation, a high-reasoning-demand task~\citep{cook2018exactly}, current work has yet to explore the thoughts (reasoning) behind the generated code. 2) Ineffective error correction. Reflection-based approaches merely accumulate historical errors in memory without providing correct reasoning pathways~\citep{zhou2023language, shinn2024reflexion}, making it difficult for subsequent search iterations to identify optimal solutions, resulting in diminished search quality.

This paper presents a novel perspective on the problem by introducing a direct thought revision approach. Previous work \citep{wang2024planning} has established that correct reasoning processes lead to correct code, and we leverage this insight to achieve accurate code generation through continuous refinement of the underlying thought processes. As illustrated in Figure~\ref{fig:intro}, traditional \textit{reflection} mechanisms merely append historical errors without actively refining the reasoning trajectory, requiring subsequent search algorithms to process increasingly lengthy memory traces. Our approach directly refines erroneous thoughts, enabling the natural emergence of correct reasoning pathways. This targeted refinement strategy significantly improves efficiency by addressing the root causes of errors rather than accumulating extensive error histories.

In light of this, we develop \our{}, a thought-search framework for code generation that simultaneously searches and refines reasoning based on code execution feedback. Specifically, \our{} begins by employing the MCTS algorithm to explore reasoning paths before generating code and then generates the code based on these reasoning thoughts. After executing the code, we perform a block-level analysis on the code and construct the verbal feedback. Following this, we introduce a refinement mechanism called \textit{rethink}, which makes the LLM refine erroneous thoughts based on the feedback. As shown in Figure~\ref{fig:intro}, this enables the search algorithm to continue exploring along corrected paths, ultimately enhancing the search tree's quality. To further guide action evaluation in the MCTS search process, we propose a dual evaluation approach to ensure effective code selection, particularly when public test cases alone are insufficient. Extensive experiments not only demonstrate the effectiveness of \our{}, but also reveal the critical factors enabling successful tree search in code generation.
Our main contributions can be summarized as follows:

\begin{itemize}[leftmargin=*]
    \item \textbf{Reasoning-to-Code Search Framework for Code Generation}: Our framework employs a multi-step thinking process combined with code generation using Monte Carlo Tree Search (MCTS) to explicitly explore various strategies for code generation. A combination of verbal and scalar feedback guides the MCTS tree generation. To the best of our knowledge, we are the first to search and refine the thought process behind code to enhance LLMs on code generation. 
    \item \textbf{Refining Erroneous Thoughts in MCTS}: We introduce the \textit{rethink} mechanism into MCTS to refine erroneous thoughts using detailed verbal feedback from code execution, allowing the search to follow higher-quality traces. Different from reflection-based methods that summarize past errors without changing current erroneous reasoning, our approach directly refines flawed thoughts, ensuring the search proceeds along more optimal trajectories.
    \item \textbf{Introducing Detailed Feedback and Dual Evaluation for Refinement}: Block-level analysis is introduced as the detailed feedback of code execution, guiding the refinement of faulty thought. Additionally, a dual evaluation method—using both public test cases and LLM self-evaluations—is used to ensure effective code selection, particularly when public test cases alone cannot fully assess the code’s correctness.
\end{itemize}

\section{Related Work}
\label{rela_work}
\paragraph{LLMs for Code Generation} Large language models (LLMs) have been widely applied and developed in the field of code~\citep{nam2024using, huang2023bias, li2024enhancing, he2024let}. Research on LLMs for code generation falls into two paradigms: (1) \textit{Code-specialized fine-tuning} that enhances syntax understanding through targeted training~\citep{luo2023wizardcoder, li2023starcoder, fried2022incoder, roziere2023code}. (2) \textit{LLM-as-agent} frameworks where models orchestrate code generation~\citep{ishibashi2024self, zhang2024codeagent, jin2024llms}. LDB proposed by ~\citet{zhong2024ldb} takes the LLM as a debugger and utilizes block-level decomposition to locate bugs. PG-TD proposed by ~\citet{zhang2023planning} utilizes Monte Carlo Tree Search ~\citep{browne2012survey} methods combined with the probabilistic output of LLMs to achieve token-level search for code generation. While effective, these approaches neglect explicit modeling of the semantic reasoning essential for complex coding tasks—a gap our work addresses.

\paragraph{Tree Search-enhanced LLMs} Tree search methods can improve the reasoning performance of LLMs by exploring various possible paths~\citep{wang2024litesearch, meng2024llm, yuan2024advancing}. By designing different action spaces, LLMs can explore at different levels~\citep{zhang2023planning, hu2024uncertainty, hao2023reasoning}. 
At the implementation level, LATS~\citep{zhou2023language} conducts code-space search while maintaining error logs as reflective memory for subsequent iterations. TS-LLM~\citep{feng2023alphazero} introduces a training-based approach with learned value functions to direct decoding trajectories.
While these methods successfully enhance the task-solving abilities of LLMs, they may not fully harness the potential of tree search in code generation tasks. This is largely because many of these approaches focus on token- or code-level searches, overlooking the deeper reasoning process that is critical for tasks like code generation, which require intricate reasoning. Additionally, the detailed execution feedback provided by the code environment has great potential to guide the search process, but these methods fall short of effectively integrating this feedback into the search.
In this paper, we focus on leveraging detailed feedback from the code execution environment to guide and refine the thought process, thereby improving the overall quality of exploration.

\section{Preliminaries}
\label{Preliminaries}
\subsection{Problem Formulation}
We focus on competition-level code generation, following the setup established by \citet{zhang2023planning}. For a given LLM, the input consists of a problem statement $P$ and a set of public test cases $T_{\text{pub}}$, each defined by an input-output pair. The goal is to develop an inference framework that enables the code generation model $M$ to produce the correct code $C \sim M(P, T_{\text{pub}})$ solving the given problem. To rigorously evaluate performance, we maintain hidden private test cases $T_{\text{priv}}$ that remain inaccessible during code generation. The primary evaluation metric is the model's ability to pass these private test cases.

\subsection{Block-level Code Analysis}
\label{blkana}
Executing buggy code in an executor can only provide standard error information. If the code runs without crashing but produces incorrect outputs, there is often little to no error feedback available. However, since code is quite structured~\citep{chevalier2007structural}, it is possible to extract detailed execution feedback through a more organized analysis.
We follow previous work by ~\citet{zhong2024ldb} to get a block-level code analysis. 

In static code analysis, the code could be divided into basic blocks~\citep{larus1999whole}. A basic block is defined as a linear sequence of code containing a single entry point and a single exit point~\citep{flow1994control, alfred2007compilers}. We first acquire the control-flow graph (CFG) of the code, and then a public test case is fed into this graph to produce an execution trace of the test, $[B_{1}, B_{2},..., B_{n}]$, where each node within the CFG corresponds to a basic block. We execute these blocks one by one and track all variable state changes in the trace. These blocks and variables are collected and then provided to the LLM to perform a block-level analysis, assessing whether each block is correct or faulty. We show an example of the analysis process in the Appendix~\ref{exm_vf}.


\begin{figure*}
    \centering
    \includegraphics[width=0.98\linewidth]{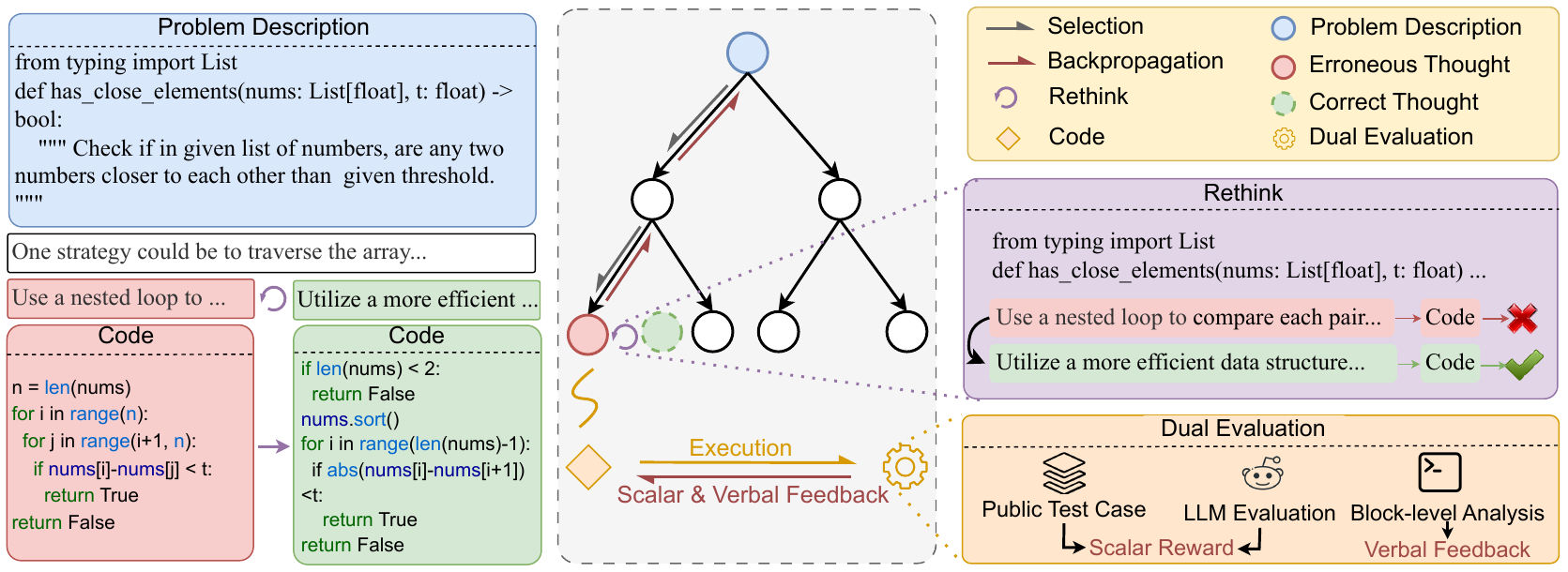}
    \caption{Overview of \our{}. We use MCTS to explore different thoughts before generating code. We obtain block-level analysis as verbal feedback through a code executor and use the verbal feedback from failed test cases to refine the thoughts, thereby improving the overall quality of the search tree.}
    \label{fig:overview}
\end{figure*}

\section{\our{}}
\label{method}
\paragraph{Overview}
\our{} is motivated by the need to search and refine the thought process during code generation using feedback from the coding environment, ultimately guiding the LLM toward correct solutions. To accomplish this, we leverage an LLM to generate both thoughts and code, iteratively refining the reasoning based on execution feedback. We employ Monte Carlo Tree Search (MCTS) as our search algorithm to optimally balance exploration and exploitation. Crucially, we introduce a novel \textit{rethink} mechanism that utilizes detailed code execution feedback to identify and refine erroneous thoughts. This approach enables the search to follow improved reasoning paths, thereby enhancing overall search quality.
The framework is shown in Figure~\ref{fig:overview}, and we provide the pseudo-code in Algorithm~\ref{alg:ouralg} in the Appendix~\ref{ag:algs}. Our design has the following key features:
\begin{itemize}[leftmargin=*]
    \item \textbf{Tree Search for Thought Process}: We employ tree search to explore the thought process of writing code. After multiple reasoning steps, code is generated based on the accumulated thoughts.
    \item \textbf{Rethink Mechanism}: We introduce a \textit{rethink} mechanism that leverages feedback from the code execution to refine and improve the quality of the reasoning process.
    \item \textbf{Block-Level Analysis Feedback}: We use block-level analysis of the code as the fine-grained feedback from code execution.
    \item \textbf{Dual Evaluation}: In our evaluation phase, we propose a dual evaluation approach, wherein both public test cases and LLM evaluation are used to assess the generated code, ultimately helping to identify high-quality solutions.
\end{itemize}

These key features are integrated into operations in \our{}, \textit{selection, expansion, evaluation, verbal feedback, backpropagation, and rethink}.

\paragraph{Selection}
In MCTS, the selection step balances exploration and exploitation by iteratively choosing the actions that are most promising for further expansion. This process continues until a leaf node is reached. Each node is selected based on a score derived from the number of visits $N(s)$ and the stored value of the state-action pair $Q(s, a)$, where the state $s$ is the problem description and prior thoughts, and action $a$ represents the new thought associated with the node. Every node's retained value $Q(s, a)$ is the maximum reward obtained by starting in $s$ and taking action $a$.
For scoring, we employ P-UCB~\citep{silver2017mastering}, an enhanced version of the UCB algorithm, to compute the overall score for each node:
\begin{multline}
\label{p_ucb}
\text{P-UCB}(s, a) = Q(s, a) \\ 
+ \beta(s) \cdot p(a \mid s) \cdot \frac{\sqrt{\log(N(s))}}{1 + N(s')} ,
\end{multline}
where $s'$ is the state reached by taking action $a$ in $s$; $N(s)$ is the visited times of the node; $p(a \mid s)$ is the probability that thought $a$ is the next thought given the problem description and previous thoughts $s$, which is proposed by the LLM agent. $\beta$ is the weight for exploration, which depends on the number of visit of $s$, defined as
\begin{equation}
\label{equ_beta}
\beta(s) = \log\left(\frac{N(s) + c_{\text{base}} + 1}{c_{\text{base}}}\right) + c ,
\end{equation}
where $c_{\text{base}}$ is a hyperparameter; $c$ is the exploration weight.  

At each state or node, the selection process chooses the action with the highest P-UCB value, and repeats this process until a leaf node is reached.

\paragraph{Expansion}
After selecting a leaf node, the expansion step generates its child nodes to explore different possible actions. We define the search action space as potential thoughts or strategies for writing the code. To make use of the feedback obtained from code execution, we handle the expansion in two scenarios: 
\begin{itemize}[leftmargin=*]
    \item If the current leaf node evaluation has failed public test cases, the expansion step incorporates the verbal feedback $f$ from these failed test cases into the prompt. The LLM then proposes multiple subsequent thoughts $z$ and assigns each thought a reasonableness score $e$, as represented by $p(a|s)$ in Eq.~(\ref{p_ucb}). The output is based on prior thoughts and the current verbal feedback, i.e., $[(z^{1}, e^{1}), \ldots ,(z^{k}, e^{k})] \sim p((z, e)^{(1 \cdots k)}|s, f)$. 
    \item If the current leaf node evaluation passes all public test cases, the expansion step directs the LLM to propose subsequent thoughts without additional feedback, i.e., $[(z^{1}, e^{1}), \ldots ,(z^{k}, e^{k})] \sim p((z, e)^{(1 \cdots k)}|s)$.
\end{itemize}
After multiple rounds of expansion, the new node's state would be the accumulated thought steps from the path to the root. We show an example of the accumulated thought steps in the Appendix~\ref{exm_accumulated_thoughts}.

\paragraph{Evaluation}
The evaluation phase in MCTS estimates the probability that a given node will successfully complete the task. While some previous works refer to this as ``simulation''~\citep{zhou2023language, hao2023reasoning}—typically involving progression from intermediate to terminal states—we evaluate nodes by generating complete code based on the current thoughts and assessing this code's quality.

In code generation, a natural evaluation is to use the pass rate of public test cases~\citep{zhang2023planning} as the reward. However, the limitation of this method is that public test cases cover only a part of the test set. When multiple code outputs pass all the public test cases, some may still fail to fully solve the problem, making it difficult to differentiate between them. To overcome this challenge, we propose a dual evaluation approach. Once all public test cases are passed, we further instruct the LLM to provide a self-assessed comprehensive score, $v^{\text{llm}}$, to evaluate the code's correctness in solving the whole problem.
\begin{equation}
  \text{reward} =
    \begin{cases}
      $$v^{\text{test}}$$, & \text{if  $0 \leq v^{\text{test}}  < 1$}\\
      $$a \times v^{\text{test}} + b \times v^{\text{llm}}$$, & \text{if $v^{\text{test}}=1$}
    \end{cases},  
    \label{reward_func}
\end{equation}
where $v^{\text{test}}$ is the pass rate on public test cases; $v^{\text{llm}}$ is the LLM's self-evaluation score. $a$ and $b$ controls the weight of two parts.

The reward in this context is a scalar value, used to calculate the Q-value at each node and to determine the score during the selection phase.
However, in code generation, the compiler and executor can return detailed error messages, and various code analysis tools can provide more granular insights into the code. These details about the code are crucial for making modifications but can not be captured in a scalar reward. Therefore, alongside the scalar reward, we also integrate verbal feedback.

\paragraph{Verbal Feedback}
When the generated code fails to pass a public test case, human programmers typically diagnose the issue by examining details such as variable values during execution. In the context of solving code generation tasks with search algorithms, relying solely on scalar feedback based on the pass rate of public test cases lacks detailed information. Therefore, we incorporate verbal feedback in the MCTS process. Specifically, as described in Sec.~\ref{blkana}, we perform block-level analysis when the code fails a public test case and store the resulting information as verbal feedback in the current node. This feedback is then utilized in both the \textit{expansion} and \textit{rethink} phases.

\paragraph{Backpropagation}
In MCTS, backpropagation refers to the process of updating the Q-values of all nodes along the path from the current node to the root node using the rewards obtained from the evaluation. Beyond using scalar feedback to update the values of parent nodes, verbal feedback is also stored in the current leaf node for use in subsequent \textit{expansion} and \textit{rethink} phases.

\paragraph{Rethink}
When the code fails to pass a public test case, we can obtain block-level analysis as detailed verbal feedback on the execution. How can we leverage such fine-grained feedback to produce correct code? We propose to use this feedback to make the LLM ``rethink'', meaning to regenerate the current erroneous thought based on the feedback to avoid generating the incorrect code. As shown in Figure~\ref{fig:overview}, the leaf node is re-generated by $z^{\text{new}} \sim p(z|s, f, z^{\text{old}})$. It is important to emphasize that we do not regenerate the parent nodes in the trace for two key reasons: 1) The parent nodes have already accumulated rewards over multiple rounds of evaluation from all their child nodes, and regenerating them would invalidate the previously gathered rewards. 2) The parent node has already gone through its own \textit{rethink} process. This means that either the parent node did not encounter failing public test cases during its evaluation or has already been refined through the \textit{rethink} process.

The advantage of introducing \textit{rethink} is twofold. From the code generation perspective, \textit{rethink} refines the reasoning process behind writing code, thus would ultimately lead to better code. From the MCTS perspective, it refines the current action or current node. Since the MCTS tree is built incrementally, improving the quality of the current action allows the LLM to explore more optimal paths in the vast search space, thereby enhancing the overall search quality of the tree. Through the \textit{rethink} mechanism, we seamlessly integrate the process of refining the reasoning of code generation with the MCTS search process.


\begin{table*}[t]
\centering
\resizebox{\linewidth}{!}
{%
\begin{tabular}{rlrrrrrrrrrr}
\toprule[1.5pt]
\multicolumn{2}{r}{\multirow{2}[4]{*}{}} & \multicolumn{4}{c}{\textbf{Pass Rate (\%)}} &  & \multicolumn{5}{c}{\textbf{Pass@1 (\%)}} \\
\cmidrule{3-6} \cmidrule{8-12}
\multicolumn{2}{r}{} & \multicolumn{1}{l}{\textbf{APPS Intro.}} & \multicolumn{1}{l}{\textbf{APPS Inter.}} & \multicolumn{1}{l}{\textbf{APPS Comp.}} & \multicolumn{1}{l}{\textbf{Average}} & & \multicolumn{1}{l}{\textbf{APPS Intro.}} & \multicolumn{1}{l}{\textbf{APPS Inter.}} & \multicolumn{1}{l}{\textbf{APPS Comp.}} & \multicolumn{1}{l}{\textbf{HumanEval}} & \multicolumn{1}{l}{\textbf{Average}} \\
\midrule
\multicolumn{1}{c}{\textbf{GPT-3.5-turbo}} & Base(1)   & 50.43 & 40.57 & 23.67 & 38.22 & & 29 & 19 & 9  & 70.12  & 37.07 \\
      & Base(16)   & 66.77 & 62.65 & 25.5 & 51.64 & & 45 & 34 & 9  & 81.71 & 47.84 \\
      & PG-TD   & 60.89 & 50.80 & 26.50 & 46.06 & & 40 & 25 & 8  & 76.22 & 42.67 \\
      & ToT   & 62.56 & 57.97 & 28.00 & 49.51 & & 38 & 25 & 10  & 76.22 & 42.67 \\
      & LATS   & 54.06 & 45.86 & 21.83 & 40.58 & & 36 & 20 & 7  & 79.88 & 41.81 \\
      & RAP   & 43.22 & 43.32 & 22.83 & 36.46 & & 21 & 14 & 8  & 71.95 & 34.69 \\
      & LDB   & 56.68 & 46.78 & 21.00 & 41.49 & & 35 & 22 & 8  & 81.09 & 42.67 \\
      & Reflexion  & 53.20 & 45.58 & 17.50 & 38.76 & & 35 & 21 & 7  & 71.95 & 39.00 \\
      \rowcolor{Gray}
      & \our{} & \textbf{67.09} & \textbf{68.65} & \textbf{29.50} & \textbf{55.08} & & \textbf{45} & \textbf{38} & \textbf{13} & \textbf{89.02} & \textbf{52.15} \\
\midrule
\multicolumn{1}{c}{\textbf{GPT-4o-mini}} & Base(1)   & 56.56 & 52.40 & 35.00 & 47.98 & & 35 & 29 & 16  & 87.20 & 48.06 \\
      & Base(16)   & 67.79 & 66.25 & 38.5 & 57.51 & & 47 & 41 & 21  & 93.29 & 56.46 \\
      & PG-TD   & 66.97 & 67.15 & 39.83 & 57.98 & & 47 & 43 & 23  & 91.46 & 56.68 \\
      & ToT   & 71.03 & 67.84 & 37.17 & 58.08 & & 52 & 46 & 23  & 92.68 & 58.84 \\
      & LATS   & 69.46 & 67.65 & 35.83 & 57.65 & & 50 & 45 & 19  & 93.29 & 57.54 \\
      & RAP   & 64.24 & 57.25 & 37.67 & 53.05 & & 39 & 32 & 20  & 87.20 & 50.43 \\
      & LDB   & 60.64 & 60.78 & 40.33 & 53.91 & & 40 & 38 & 23  & 90.85 & 53.87 \\
      & Reflexion   & 60.65 & 56.87 & 38.00 & 51.84 & & 40 & 31 & 18  & 90.85 & 51.29 \\
      \rowcolor{Gray}
      & \our{} & \textbf{76.60} & \textbf{74.35} & \textbf{42.50} & \textbf{64.48} & & \textbf{59} & \textbf{49} & \textbf{28} & \textbf{94.51} & \textbf{62.93} \\
\bottomrule[1.5pt]
\end{tabular}%
}
\caption{Performances of \our{} and baselines on APPS and HumanEval. \our{} achieves the best performance across all the datasets with the maximum number of rollouts for tree search algorithms being 16.
}
\vspace{-10pt}
\label{tab:main}
\end{table*}%

\section{Experiment Settings}
\label{experiments}
\paragraph{Datasets} 
We evaluate \our{} and baseline methods on two widely used benchmark datasets: APPS~\citep{hendrycks2021measuring} and HumanEval~\citep{chen2021evaluating}. The APPS dataset is a huge dataset contains three levels of difficulties: introductory, interview, and competition.
Within each difficulty, the problems are randomly distributed. Therefore, we elected the first 100 problems per difficulty to maintain randomness while ensuring balanced coverage, which mirrors sampling methods used by ~\citet{zhang2023planning}. 
We use \textit{pass rate} and \textit{pass@1} as the evaluation metrics for code correctness following ~\cite{zhang2023planning}. \textit{Pass rate} is the average percentage of private test cases successfully passed by the generated code across all problems, and \textit{pass@1} measures the percentage of problems where the generated programs pass all private test cases~\citep{austin2021program, chen2021evaluating, dong2023codescore}.

\paragraph{Baselines} To illustrate the effectiveness of \our{}, we compare two kinds of code generation methods. The first kind is feedback-enhanced, which uses the code execution feedback to refine codes iteratively: LDB~\citep{zhong2024ldb}, Reflexion~\citep{shinn2024reflexion}. The second kind is tree search-enhanced methods:  PG-TD~\citep{zhang2023planning}, ToT~\citep{yao2024tree}, LATS~\citep{zhou2023language} and RAP~\citep{hao2023reasoning}. More details can be found in Appendix ~\ref{ad:bsls}.

\paragraph{Implementation}
We pick GPT-3.5-turbo and GPT-4o-mini as the backbone models to compare different algorithms. For search-enhanced methods, including \our{}, we set the maximum number of children of any node to be 3. For MCTS-based methods, we set the hyperparameters in Eq.~(\ref{equ_beta}) $c_{\text{base}}$ to be 10 and $c$ to be 4 following previous work by ~\citet{zhang2023planning}. And we set the $a$ and $b$ in Eq.~(\ref{reward_func}) to be $(0.8, 0.2)$ and we compare performances under different settings in Sec.~\ref{exp_res}. We set the maximum number of rollouts or simulation times to be 16. For LDB, we set the maximum number of debug times to be 10, as in the original paper ~\citep{zhong2024ldb}.

\section{Results And Analysis}
\label{exp_res}
\paragraph{Overall Performance} We present the overall performance in Table~\ref{tab:main}, where we can see that \our{} outperforms all baseline models across both datasets. Additionally, by comparing them with the original base model, both feedback-enhanced and tree search-enhanced methods show significant performance improvements, demonstrating the effectiveness of exploring different strategies and using detailed feedback from code execution. Generally, \our{} enhances performance more significantly on GPT-3.5-turbo than on GPT-4o-mini. This may be because weaker code models benefit more from error correction in the thought process.

\begin{figure}[t]
    \centering
    \begin{minipage}{0.48\textwidth}
        \centering
        \includegraphics[width=0.98\textwidth]{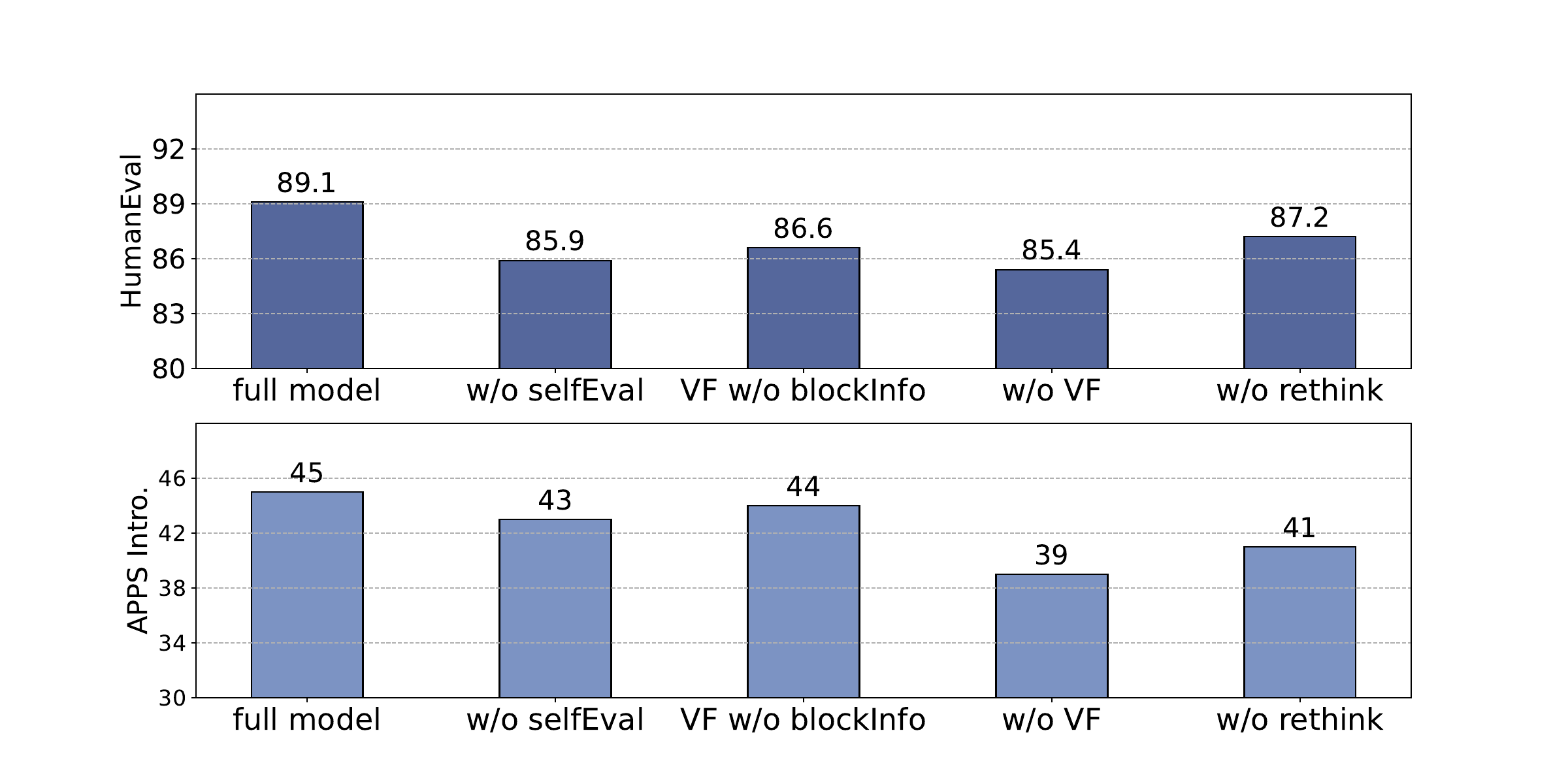}
        \caption{Ablation study of block-level analysis (blockInfo), rethink mechanism, verbal feedback (VF), and self-evaluation with GPT-3.5-turbo as the backbone.}
        \label{fig:ablation}
    \end{minipage} \hfill
    \begin{minipage}{0.48\textwidth}
        \centering
        \includegraphics[width=0.98\textwidth]{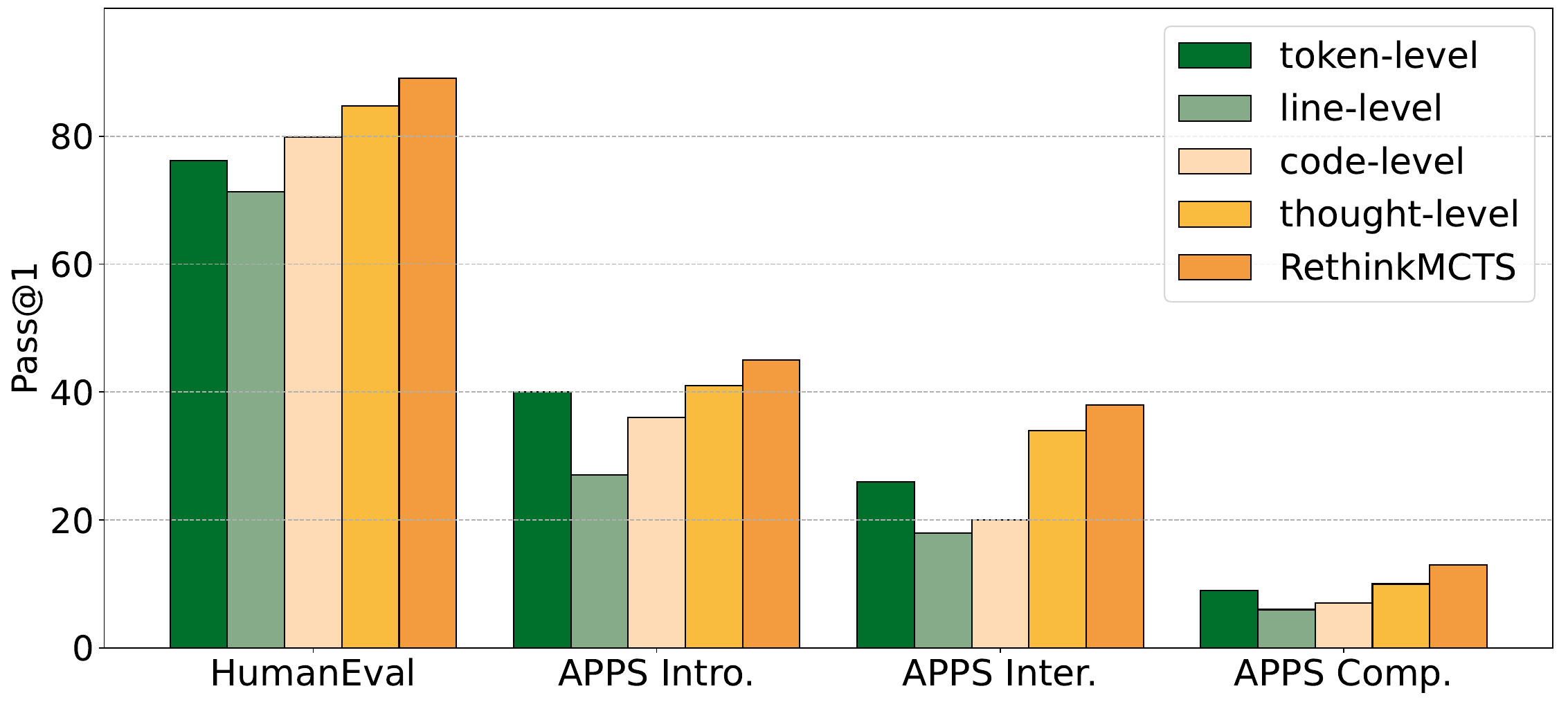}
        \caption{Performance comparison between different search granularity. For advanced models like GPT-3.5-turbo, it's better to explore at the thought level.}
        \label{fig:action_level}
    \end{minipage}
    \vspace{-12pt}
\end{figure}

\paragraph{Ablation Study} We conduct ablation studies to remove each of our model's components, including self-evaluation (w/o selfEval), block-level analysis (partial verbal feedback, w/o blockInfo), whole verbal feedback (w/o VF), rethink mechanism (w/o \textit{rethink}). The results using GPT-3.5-turbo as the backbone model are shown in Figure~\ref{fig:ablation}, and the results on GPT-4o-mini are presented in Appendix~\ref{ad:results}.
The results demonstrate that each component contributes to overall performance, with verbal feedback showing the most significant impact. This aligns with our design, as the \textit{rethink} mechanism depends primarily on execution feedback—without this feedback, the model lacks the necessary information to refine its reasoning effectively.

Additionally, we can see that for the HumanEval dataset, block-level analysis significantly impacts performance  (89.1 $\xrightarrow{}$ 86.6), while its effect on APPS is minimal. We attribute this to HumanEval having fewer public test cases than APPS (2.8 vs. 27.52 on average), making detailed test case analysis essential for the \textit{rethink} mechanism to correct errors in HumanEval. This explains why dual evaluation is critical for HumanEval - the limited test cases necessitate LLM-based code reevaluation. Finally, the \textit{rethink} mechanism we proposed significantly enhances the results. This improvement stems from that \textit{rethink} enabling the use of fine-grained block-level analysis in verbal feedback, effectively correcting logical errors in the reasoning process.

\paragraph{Search Granularity Study}
\our{} conducts a thought-level search for code. Here, we compare the action spaces for MCTS, specifically examining 4 levels of search granularity: token, line, code, and thought. The experimental results with GPT-3.5-turbo as backbone are presented in Figure~\ref{fig:action_level}, and the results on GPT-4o-mini are presented in Appendix~\ref{ad:results}. 

As shown in the figure, the thought-level search is more effective in finding viable code. This demonstrates that for advanced LLMs like GPT-3.5-turbo, exploring the reasoning process is beneficial~\citep{zhang2024llm, huang2022towards}. Additionally, we observe that token-level searching performs better than line and code-level searching. This is due to the fact that with a limited number of search iterations, token-level searches allow fewer constraints on the early tokens, thus uncovering more possibilities compared to line and code-level searches. Finally, although thought-level search yields the best results among different granularity, its effectiveness is further enhanced in \our{} by introducing detailed feedback and \textit{rethink} mechanism, making the search over thoughts in the code generation process even more effective.

\paragraph{Rethink vs. Reflection}
In this section, we compare \textit{rethink} and \textit{reflection} approaches. Our comparison methodology maintains all other components of \our{} unchanged, with the only difference being the replacement of \textit{rethink} with \textit{reflection}. The experimental results are presented in Table \ref{tab:rethink_vs_reflection}. As demonstrated, \textit{rethink} not only improves search effectiveness compared to \textit{reflection} but also significantly reduces token consumption. This efficiency stems from \textit{rethink}'s ability to directly modify incorrect thought steps, whereas \textit{reflection} continuously accumulates error history, leading to excessive token consumption. Furthermore, since erroneous steps remain in the search tree with \textit{reflection}, subsequent searches may continue down incorrect paths, resulting in inferior search performance.

\begin{table}[th!]
    \centering
    \setlength{\tabcolsep}{3pt}
    \resizebox{0.49\textwidth}{!}{
    \begin{tabular}{c|cc|cc}
        \toprule
        \multirow{2}{*}{Dataset} & Reflection & Rethink & Reflection & Rethink \\
        & \multicolumn{2}{c|}{(Pass@1)} & \multicolumn{2}{c}{(Avg. Token Cost)}  \\
         \midrule
        APPS-Intro. & 54 & \bf 59\bm{$(\uparrow 9.2\%)$} & 177353 & \bf 143048\bm{$(\downarrow 19.3\%)$}  \\
        APPS-Inter. & 45 & \bf 49\bm{$(\uparrow 8.9\%)$} & 163494 & \bf 126648\bm{$(\downarrow 22.5\%)$} \\
        APPS-Comp. & 24 &  \bf 28\bm{$(\uparrow 16.6\%)$} & 189215 & \bf 182193\bm{$(\downarrow 3.7\%)$} \\
        HumanEval & 93.29 &  \bf 94.51\bm{$(\uparrow 1.3\%)$} & 57027 & \bf 36678\bm{$(\downarrow 35.7\%)$} \\
        \bottomrule
    \end{tabular}
     }
    \caption{Comparison between rethink and reflection-based MCTS approaches. We experiment on \our{} with other parts remain the same and only replace \textit{rethink} with \textit{reflection}.}
    \vspace{-10pt}
    \label{tab:rethink_vs_reflection}
\end{table}

\begin{table*}[tb]
\centering
\resizebox{0.95\textwidth}{!}
{
\begin{tabular}{lrrrrrrrr}
\toprule[1.5pt]
\multirow{2}{*}{$\boldsymbol{(a, b)}$}

& \multicolumn{3}{c}{\textbf{Pass Rate (\%)}} & \multicolumn{1}{c}{} & \multicolumn{4}{c}{\textbf{Pass@1 (\%)}}  \\
\cmidrule{2-4} \cmidrule{6-9}\
& \multicolumn{1}{l}{\textbf{APPS Intro.}} & \multicolumn{1}{l}{\textbf{APPS Inter.}} & \multicolumn{1}{l}{\textbf{APPS Comp.}} & & \multicolumn{1}{l}{\textbf{APPS Intro.}} & \multicolumn{1}{l}{\textbf{APPS Inter.}} & \multicolumn{1}{l}{\textbf{APPS Comp.}} & \multicolumn{1}{l}{\textbf{HumanEval}} \\
\midrule[0.75pt]
(0.8, 0.2) & 76.6 & 74.3 & 42.5 &  & 59      & 49      & \textbf{28}     & \textbf{94.5} \\
(1.0, 0.2) & 76.9 & \textbf{76.4} & \textbf{43.5} &  & \textbf{60}      & 53      & 27     & 92.7 \\
(1.0, 1.0) & \textbf{78.8} & 75.2 & 40.5 &  & \textbf{60}      & \textbf{54}      & 24     & 91.5 \\
\bottomrule[1.5pt]
\end{tabular}
}
\caption{Performance comparison under different reward weights. The (1.0, 0.2) and (1.0, 1.0) configurations make the nodes that achieve a pass rate of 1.0 on public test cases receive a score higher than 1.0, whereas the (0.8, 0.2) configuration keeps all node evaluations between 0$\sim$1.}
\vspace{-3mm}
\label{tab:reward_weights}
\end{table*}

\begin{figure}[h]
    \centering
 
    \includegraphics[width=0.45\textwidth]{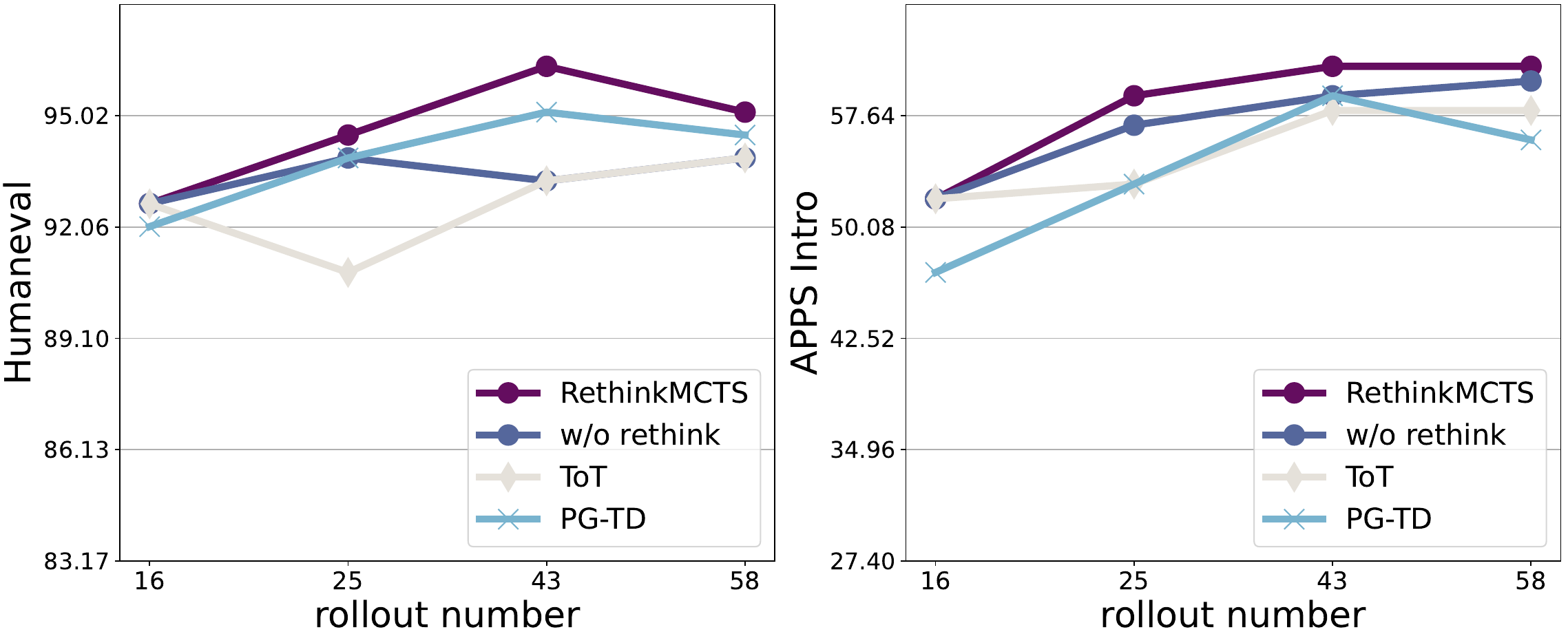}
    \caption{Performance comparison between \textit{rethink} more times and more rollouts without \textit{rethink}. \textit{rethink} is more effective than increasing rollouts.
    }
    \vspace{-12pt}
    \label{fig:rethinkMore}
\end{figure}

\paragraph{Test Time Scaling with \textit{Rethink}}
The goal of \textit{rethink} is to improve the search quality within the same number of rollouts. To validate the effectiveness of \textit{rethink}, we compare the performance between increasing the number of \textit{rethink} operations and increasing the number of rollouts without applying \textit{rethink}, while keeping the total number of rollouts consistent. The results are shown in Figure~\ref{fig:rethinkMore}.

The figure shows that increasing the number of \textit{rethink} operations and increasing the number of rollouts both enhance performance. This is expected as more extensive exploration raises the probability of finding the correct code.  However, increasing the number of \textit{rethink} operations yields greater performance gains. This can be attributed to two key reasons. From a tree search perspective, without the \textit{rethink} mechanism, erroneous actions or nodes would persist in the trace, causing the following nodes to follow incorrect reasoning paths, which makes it challenging to ensure the quality of the entire reasoning trace. From the code generation perspective, the \textit{rethink} mechanism refines flawed thoughts and get a better thought chain, which would finally lead to better codes.

\begin{table}[h]
    \centering
    \resizebox{0.9\columnwidth}{!}{
    \begin{tabular}{lcc}
         \toprule[1.5pt]
         \multirow{1}{*}{\textbf{Method}} & \multicolumn{1}{c}{\textbf{APPS Intro.}} & \multicolumn{1}{c}{\textbf{HumanEval}} \\
         \midrule[0.75pt]
         \textsc{w/o Rethink} & 10.04 & 48.30 \\
         \textsc{\our{}} & \textbf{15.60} & \textbf{53.29}  \\ 
         \bottomrule[1.5pt]
    \end{tabular}
    }
    \caption{The success rate comparison of the searched codes between with and without the \textit{rethink} mechanism.}
    \vspace{-10pt}
    \label{tab:wikipedia_only}
\end{table}

Furthermore, we compare the pass rate on public test cases of all the generated codes for the entire tree, with and without the \textit{rethink} operation, since only public test cases are available during the search. The results are presented in Table~\ref{tab:wikipedia_only}. We can see that the \textit{rethink} operation increases the proportion of effective code found in the tree. This highlights how refining erroneous thoughts enables the tree to focus more on correct paths, leading to better outcomes.

\paragraph{Study on Reward Weights} 
\label{reward_weights}
We analyzed the impact of reward weights in Eq.~(\ref{reward_func}) of Sec.~\ref{method}, with results shown in Table~\ref{tab:reward_weights}. It is evident that $(a, b)$ significantly influences \our{}'s performance, underscoring the importance of LLM self-evaluation. Since self-evaluation rewards apply only when code achieves a perfect pass rate on public test cases, each configuration yields distinct implications.

Under the (0.8, 0.2) configuration, the code is given a baseline score of 0.8, and the LLM's evaluation score is used to distinguish between different codes. This allows for situations where the total score of code that passes all public test cases could be lower than that of code with a pass rate below 1, but only when the LLM's self-evaluation score is particularly low. 
Conversely, configurations (1.0, 0.2) and (1.0, 1.0) ensure that code passing all public tests always receives a score exceeding 1.0. While this approach guarantees that final output maintains perfect public test performance, it prematurely discards promising reasoning paths with imperfect test results. This limitation explains the poorer performance observed on both datasets under these configurations.

\section{Conclusion}
We propose \our{}, the first framework that searches and refines thoughts for code generation. Unlike previous tree search methods, \our{} explores the reasoning process and incorporates an iterative \textit{rethink} mechanism to improve search quality. Compared to traditional \textit{reflection}, \textit{rethink} achieves superior results with lower token cost by guiding search along correct paths. Experiments on APPS and HumanEval datasets demonstrate that \our{} outperforms existing approaches, generating high-quality code through search-and-refinement reasoning. Beyond code generation, \our{} offers a general approach for enhancing task performance through structured reasoning, with potential applications in other LLM domains such as mathematical problem-solving and tool usage scenarios.

\section*{Acknowledgement}
The Shanghai Jiao Tong University team is partially supported by National Key R\&D Program of China (2022ZD0114804), Shanghai Municipal Science and Technology Major Project (2021SHZDZX0102) and National Natural Science Foundation of China (62322603, 62177033).

\section*{Limitations}
\paragraph{Limited Exploration of Fine-Tuning based on Collected Data}  \our{} framework generates high-quality data that includes thought steps, execution feedback, and code. This data could potentially be used to fine-tune an LLM to enhance its code generation capabilities. However, since our work focuses primarily on the inference framework, we leave the fine-tuning exploration for future work.

\paragraph{Generalization to Other Reasoning Tasks}  Our primary contribution lies in developing a search framework that integrates code execution feedback for refinement. While this approach is effective for code generation tasks, it may not generalize well to other reasoning domains, such as mathematical reasoning, where similarly detailed feedback mechanisms might not be available. Nevertheless, our method could potentially be applicable to reasoning tasks that involve detailed feedback mechanisms comparable to those in code generation.

\paragraph{Potential Limitations in the Refinement Step} Our current framework enhances the original MCTS approach by introducing a refinement step that utilizes detailed feedback from code execution. However, we do not introduce a sophisticated procedure for determining which specific thought step should be refined. Instead, we directly select the most recent step that caused the error, since in MCTS, the tree develops incrementally, ensuring each step eventually has the opportunity to be refined if it produces incorrect code. Although this approach proves effective, integrating a dedicated verifier to identify which thought step requires refinement could potentially yield better results.

\section*{Ethics Statement}
In this work, we employ LLMs as both thought and code generators. All the dataset we use are publicly available and are for research purposes only. The LLMs utilized in our study include the closed-source model GPT-4o-mini and GPT-3.5-Turbo. Ethical considerations related to these models, including their training data and deployment, are addressed by their respective creators. The LLMs in our work are instructed solely to output code task-related thought steps, evaluation scores and code, and do not generate other free-form text. However, we acknowledge that LLMs, including those used in our study, may occasionally produce improper or harmful content. Such outputs are unintended and do not reflect the views or intentions of the authors.


\bibliography{custom}


\appendix
\section*{Appendix}
\section{Details of Baselines}
\label{ad:bsls}
Here, we present more details of the implementation of the baselines:

(1) Code Generation Algorithms:
\begin{itemize}[topsep = 3pt,leftmargin =15pt]
    \item \textbf{LDB}~\citep{zhong2024ldb}: A debugging framework that divides the initial code into blocks, analyzes each block, and resolves issues by monitoring changes in block-level variable values. It iteratively optimizes the code by following this process.
    \item \textbf{Reflexion}~\citep{shinn2024reflexion}:Iteratively refine the initial code by utilizing historical error data and incorporating insights gained from previous errors.
\end{itemize}

(2) Tree Search-enhanced Methods:
\begin{itemize}[topsep = 3pt,leftmargin =15pt]
    \item \textbf{PG-TD}~\citep{zhang2023planning}: A token-level MCTS search method that uses the code's pass rate as a scalar reward.
    \item \textbf{ToT}~\citep{yao2024tree}: We apply the Tree-of-Thoughts (ToT) approach to code generation in a manner similar to its application in creative writing. The search process is structured into two distinct phases: thought generation and code generation, with the tree being explored using a breadth-first search (BFS) strategy.
    \item \textbf{LATS}~\citep{zhou2023language}: A framework that integrates MCTS with reflection, summarizing past errors and storing them as memory within nodes to assist with future iterations.
    \item \textbf{RAP}~\citep{hao2023reasoning}: Leveraging an LLM as the world model to simulate and evaluate search results.
\end{itemize}

\section{Additional results}
\label{ad:results} %
This section presents some additional experiment results. 

\begin{figure}[ht]
    \centering
    \begin{minipage}{0.48\textwidth}
        \centering
        \includegraphics[width=0.98\textwidth]{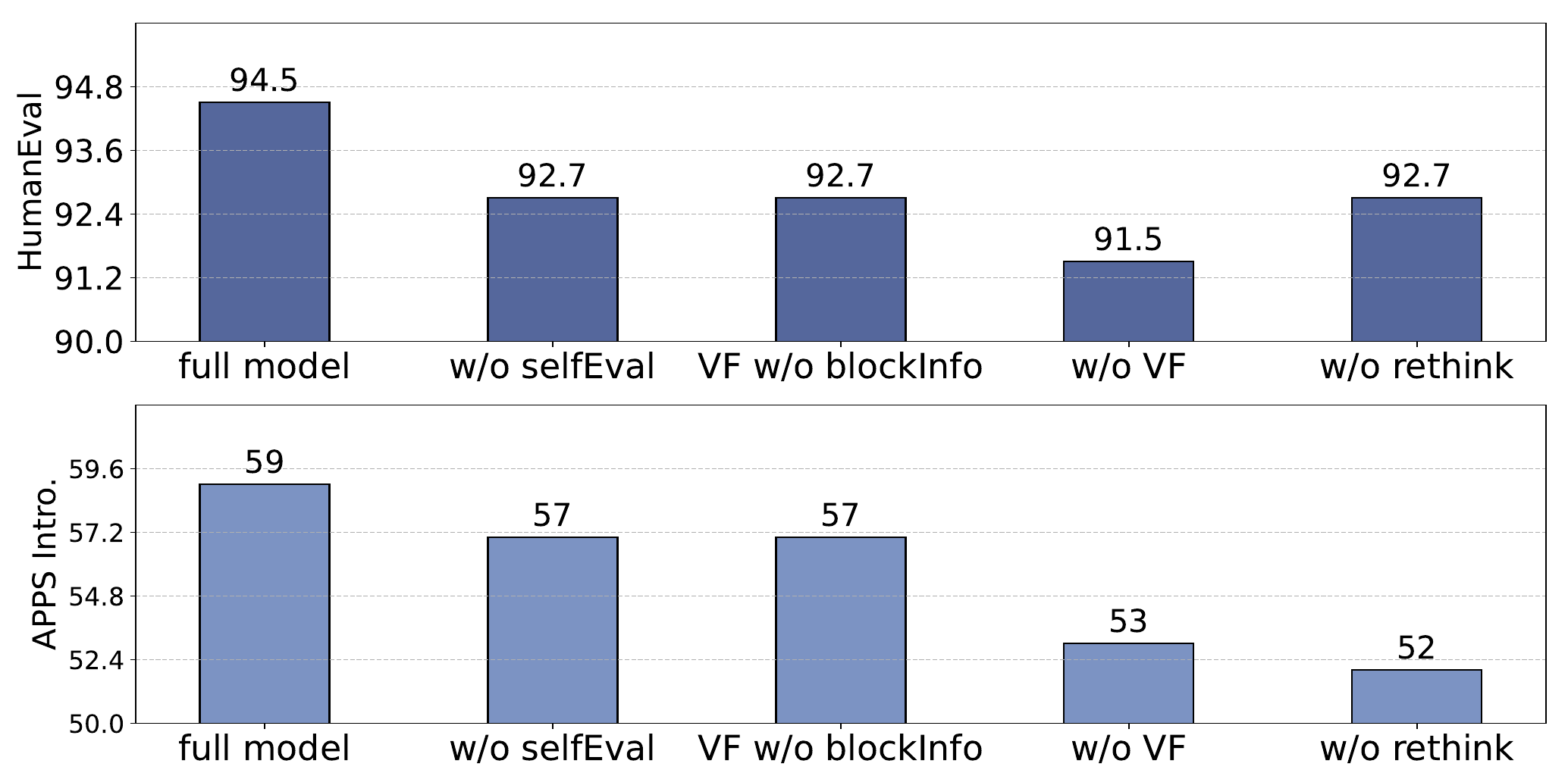}
        \caption{Ablation study of block-level analysis (blockInfo), rethink mechanism, the verbal feedback (VF) and self-evaluation with GPT-4o-mini as the backbone.}
        \label{fig:ablation_mini}
    \end{minipage} \hfill
    \begin{minipage}{0.48\textwidth}
        \centering
        \includegraphics[width=0.98\textwidth]{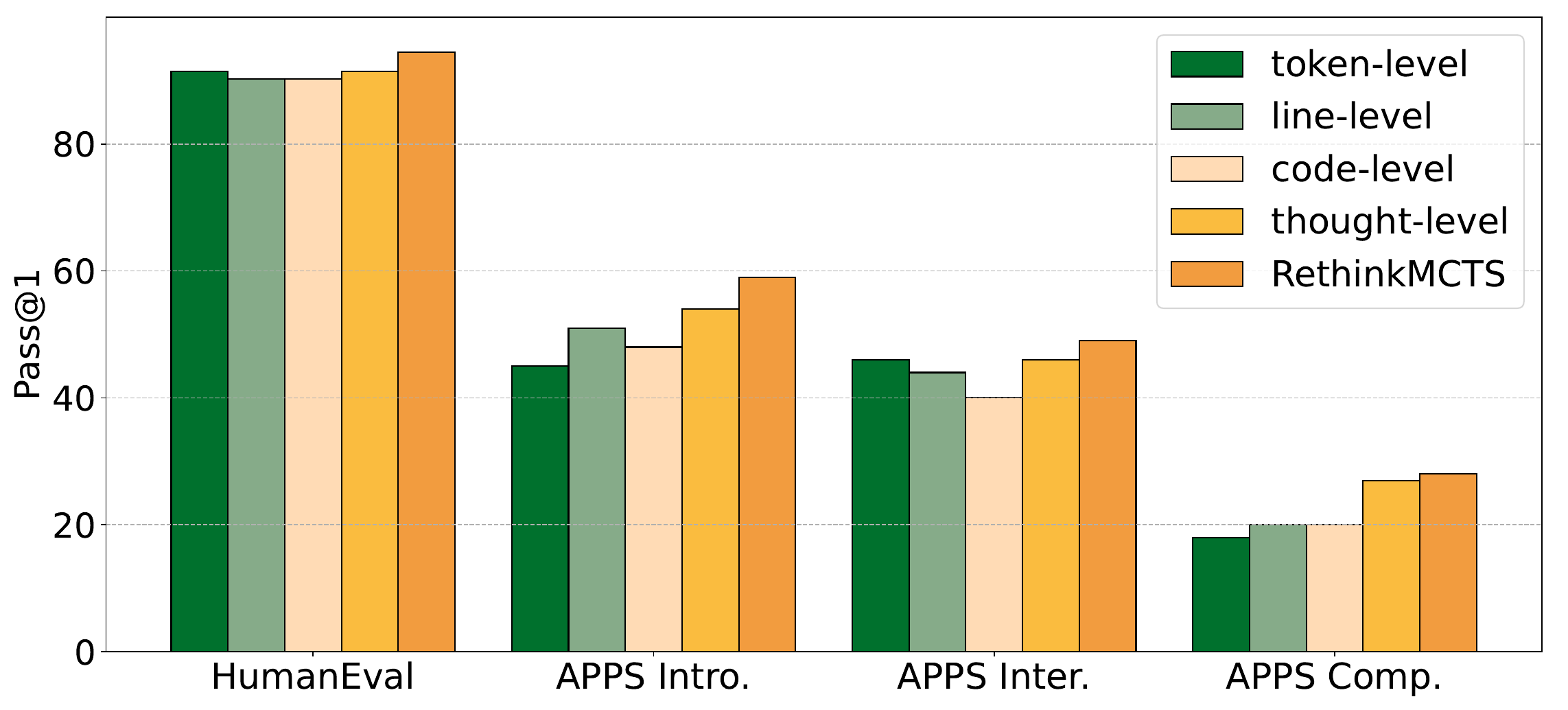}
        \caption{Performance comparison between different search granularity. For advanced model like GPT-4o-mini, it's better to explore at the thought level.}
        \label{fig:action_level_mini}
    \end{minipage}
\end{figure}

\paragraph{Ablation Study} Here, we present the results of the ablation study using GPT-4o-mini as the backbone model, as shown in Figure~\ref{fig:ablation_mini}.  It is clear that the \textit{rethink} operation and verbal feedback remain the most significant contributors to our model’s performance. Notably, the \textit{rethink} mechanism exhibits even stronger effects with GPT-4o-mini than with GPT-3.5-turbo, likely due to the model’s enhanced ability to effectively utilize feedback and make refinement.

\paragraph{Search Granularity Study}
We present the results of the search granularity study using GPT-4o-mini as the backbone model, shown in Figure~\ref{fig:action_level_mini}. It is evident that the differences across granularities are smaller on the HumanEval dataset, likely due to its relatively low overall difficulty. However, on the APPS dataset, the advantage of thought-level search becomes much more pronounced, especially at the highest ``competition'' difficulty level. This suggests that for more complex problems, exploring the thought process and reasoning is beneficial.

\paragraph{Token Consumption}
Our approach ensures the comparison fair by keeping the number of rollouts, i.e., the number of codes generated during the search, the same. This is following the previous work about tree search in code generation PG-TD~\citep{zhang2023planning} and LATS~\citep{zhou2023language}. 

However, one limitation is that our method would cost more tokens since we have introduce block-level analysis and rethink mechanism. Here we present the detailed token usage on GPT-4o-mini in Table~\ref{tab:tk_num}.

Our increase in token usage is primarily due to the introduction of block-level analysis, which includes the values of variables before and after each block. These values are obtained through code execution, resulting in longer textual inputs to the LLM. However, the feedback makes our model deliver significantly better results. It is essential to the success of the "rethink" mechanism, and it represents an important characteristic in the coding environment (there would be no such detailed feedback for the math reasoning problem). Therefore, incorporating such feedback is crucial. Like OpenAI's o1 model (which solves one problem with thousands of tokens in the hidden CoT and minutes to take), our primary aim is not to optimize token count or computation time. Instead, the emphasis is on enabling LLMs to generate higher-quality reasoning processes and achieve superior reasoning outcomes.
\begin{table*}
\centering
\resizebox{0.85\linewidth}{!}
{
    \begin{tabular}{lrrrrrrr}
    \toprule[1.5pt]
    & \multicolumn{3}{c}{\textbf{APPS Intro. (\%)}} & & \multicolumn{3}{c}{\textbf{HumanEval}}\\
    \cmidrule{2-4} \cmidrule{6-8}\
    & \multicolumn{1}{l}{\textbf{\#Input token}} & \multicolumn{1}{l}{\textbf{\#Output token}} & \multicolumn{1}{l}{\textbf{Cost(\$)}}&  & \multicolumn{1}{l}{\textbf{\#Input token}} & \multicolumn{1}{l}{\textbf{\#Output token}} & \multicolumn{1}{l}{\textbf{Cost(\$)}} \\
    \midrule[0.75pt]
    ToT & 24799 & 7156 & 0.008 & & 11687 & 7131 & 0.006 \\
    LATS & 104634 & 17472 & 0.026 & & 12690 & 7403 & 0.006 \\
    PG-TD & 27827 & 5378 & 0.007 & & 5959 & 3759 & 0.003 \\
    LDB & 61112 & 1734 & 0.010 & & 13161 & 480 & 0.002 \\
    \our{} & 123207 & 17863 & 0.029 & & 28479 & 8198 & 0.009 \\ 
    \bottomrule[1.5pt]
    \end{tabular}
}
\caption{The token consumption comparison. The results represent the average number of tokens consumed per question.}

\label{tab:tk_num}
\end{table*}

\paragraph{Self-evaluation vs. Self-generating Unit Tests}
Given the limited coverage of public test cases, we propose a dual evaluation approach. In this section, we compare it with an alternative approach of self-generating unit tests. In the latter approach, when the code passes the public test cases, we have the LLM generate additional test cases and get a new \textit{pass rate} on these tests. The combined results serve as a comprehensive evaluation of the code. Experimental results are shown in Table~\ref{tab:lm_vs_gentests}. 

As the table demonstrates, while self-generating unit tests improve the \textit{pass rate} on test cases, they do not improve the \textit{pass@1} metric. This is because self-evaluation directly assesses the code after it passes the public test cases, scoring it based on how well it meets the problem's requirements. As a result, it provides a more accurate indication of the code's ability to address the entire problem. In contrast, self-generating unit tests focus on creating additional tests, which emphasize the test suite rather than the code itself. There are two potential reasons for this: 1) Self-generating unit tests primarily identify patterns in the existing tests and generate a set of tests that better match the test suite. This can enhance the \textit{pass rate} by filtering for code that matches these patterns, but it doesn't necessarily identify the mismatch between the code and the problem requirement. 2) The generated tests may not always be correct~\citep{huang2023agentcoder}, which can mislead the code's modification process and the subsequent search direction, potentially steering it away from valid solutions.

\begin{table*}
\centering
\resizebox{\linewidth}{!}
{
    \begin{tabular}{lrrrrrrrr}
    \toprule[1.5pt]
    & \multicolumn{3}{c}{\textbf{Pass Rate (\%)}} & & \multicolumn{4}{c}{\textbf{Pass@1 (\%)}}\\
    \cmidrule{2-4} \cmidrule{6-9}\
    & \multicolumn{1}{l}{\textbf{APPS Intro.}} & \multicolumn{1}{l}{\textbf{APPS Inter.}} & \multicolumn{1}{l}{\textbf{APPS Comp.}}&  & \multicolumn{1}{l}{\textbf{APPS Intro.}} & \multicolumn{1}{l}{\textbf{APPS Inter.}} & \multicolumn{1}{l}{\textbf{APPS Comp.}} & \multicolumn{1}{l}{\textbf{HumanEval}} \\
    \midrule[0.75pt]
    Direct Evaluation  & 76.60  & 74.34 & 42.50 &  & 59 & 49 & 28 & 94.51 \\
    Self-generated Tests& 77.32 & 75.80 & 47.23 & & 59 & 44 & 28 & 93.29  \\   
    \bottomrule[1.5pt]
    \end{tabular}
}
\caption{The performance comparison between using Direct Self-evaluation and Self-generating test evaluation. }

\label{tab:lm_vs_gentests}
\end{table*}

\paragraph{Multiple Runs}
To further illustrate our method's advantage, we run our method and strong baselines for 3 times with different random seeds. The average performance and the standard derivation are presented in Table~\ref{tab:multi_runs}. Noticeably, our model, \our{}, consistently demonstrates a stable performance advantage across multiple experiments. Additionally, we have noticed that since we set the temperature to 0, the standard deviation between different runs is very small. Therefore, we included the result of only one run in the main body of the text.

\begin{table*}
\centering
\resizebox{\linewidth}{!}
{
    \begin{tabular}{lrrrrrrrr}
    \toprule[1.5pt]
    & \multicolumn{3}{c}{\textbf{Pass Rate (\%)}} & & \multicolumn{4}{c}{\textbf{Pass@1 (\%)}}\\
    \cmidrule{2-4} \cmidrule{6-9}\
    & \multicolumn{1}{l}{\textbf{APPS Intro.}} & \multicolumn{1}{l}{\textbf{APPS Inter.}} & \multicolumn{1}{l}{\textbf{APPS Comp.}}&  & \multicolumn{1}{l}{\textbf{APPS Intro.}} & \multicolumn{1}{l}{\textbf{APPS Inter.}} & \multicolumn{1}{l}{\textbf{APPS Comp.}} & \multicolumn{1}{l}{\textbf{HumanEval}} \\
    \midrule[0.75pt]
    ToT      & 72.22$\pm$1.19  & 67.10$\pm$1.06  & 40.33$\pm$2.58 &   & 53.67$\pm$1.70 & 45.00$\pm$0.82 & 22.67$\pm$2.05  & 92.48$\pm$0.29  \\
    LATS     & 70.17$\pm$0.52  & 68.66$\pm$0.79  & 36.83$\pm$3.61 &  & 50.33$\pm$0.47 & 45.67$\pm$0.94 & 18.33$\pm$3.30  & 93.70$\pm$0.57  \\
    PG-TD    & 68.85$\pm$2.29  & 68.29$\pm$1.48  & 40.33$\pm$2.00 &  & 49.00$\pm$4.32 & 44.33$\pm$1.25 & 23.00$\pm$1.63 & 92.28$\pm$0.76 \\
    \our{} & \textbf{75.36}$\pm$1.08 & \textbf{74.10}$\pm$0.98 & \textbf{43.33}$\pm$1.18 & & \textbf{57.33}$\pm$1.25 & \textbf{50.00}$\pm$0.82 & \textbf{27.00}$\pm$0.82 & \textbf{94.31}$\pm$0.29 \\
    \bottomrule[1.5pt]
    \end{tabular}
}
\caption{Comparing \our{} with competitive baselines in multiple runs. The mean and standard deviation of the results are presented.}

\label{tab:multi_runs}
\end{table*}

\section{Prompts}
\label{pt:prompts} %
In this section, we present the prompts used when an LLM to perform various operations.

\subsection{Expansion Prompt}
First, we discuss the prompts for the Expansion step in the MCTS process. There are two sets of prompts: one set is used to generate new thoughts based on the problem description and previous thoughts when there is no feedback presented in Table ~\ref{tab:prompt_gen_thoughts}; 

\begin{table*}[h!]
\centering
\begin{tcolorbox}[
    colback=blue!5!white,
    colframe=black!55!black,
    width=0.98\textwidth,
    title={Prompt for Rethink},
    listing engine=listings,
    boxrule=0.5pt
]
\small
\begin{lstlisting}
{problem statement}
{thoughts}

Above is a problem to be solved by Python program.

* I need you to analyze and provide new thoughts that can lead to the correct solution code.
* If there are previous thoughts provided, please follow them and offer more detailed and further insights, as a detailed thinking or enhancement for previous ones.
* I need you to output \{width\} possible thoughts. Remember each only contain one possible distinct reasoning but all following previous thoughts if there are.
* Please wrap your response into a JSON object that contains keys `Thought-i` with i as the number of your thought, and key `Reasonableness` with the Reasonableness of each thought, which should between 0~1 and the sum should be 1.
* The JSON should be a **list of dicts**, the dicts are split with comma ','.
Example Answers:
[
{"Thought-1":" We could use the print function to finish the task in one line: print(2 + 3)", "Reasonableness": 0.7},
{"Thought-2":" We should calculate the problem by setting a=2+3, and then print(a)", "Reasonableness": 0.29},
{"Thought-3":" The problem can't be solved by Python.", "Reasonableness": 0.01}
]
\end{lstlisting}
\end{tcolorbox}
\caption{Prompt for generating thoughts in search methods.}
\label{tab:prompt_gen_thoughts}
\end{table*}

The other set is used when the generated code contains errors and verbal feedback is provided. In this case, the LLM uses the verbal feedback to generate thoughts that avoid such errors. We present the prompt in Table~\ref{tab:prompt_gen_thoughts_vb}.

\begin{table*}[h!]
\centering
\begin{tcolorbox}[
    colback=blue!5!white,
    colframe=black!55!black,
    width=0.98\textwidth,
    title={Prompt for Rethink},
    listing engine=listings,
    boxrule=0.5pt
]
\small
\begin{lstlisting}
{problem statement}

{thoughts}
```python
generated code
```
{verbal feedback}
Above is a problem to be solved by Python program.

* I need you to analyze and provide new thoughts that can lead to the correct solution code.
* The goal is that the thoughts could lead to the code that not only avoids the current error but also solve the problem in a way that handles other potential test cases that we haven't encountered yet.
* I need you to output \{width\} possible thoughts. Remember each only contain one possible distinct reasoning but all following previous thoughts if there are.
* Please wrap your response into a JSON object that contains keys `Thought-i` with i as the number of your thought, and key `Reasonableness` with the Reasonableness of each thought, which should between 0~1 and the sum should be 1.
* The JSON should be a **list of dicts**, the dicts are split with comma ','.
Example Answers:
[
{"Thought-1":" We could use the print function to finish the task in one line: print(2 + 3)", "Reasonableness": 0.7},
{"Thought-2":" We should calculate the problem by setting a=2+3, and then print(a)", "Reasonableness": 0.29},
{"Thought-3":" The problem can't be solved by Python.", "Reasonableness": 0.01}
]
\end{lstlisting}
\end{tcolorbox}
\caption{Prompt for generating thoughts in search methods.}
\label{tab:prompt_gen_thoughts_vb}
\end{table*}

\subsection{Code Generation Prompt}
We present the prompt we use to instruct the LLM to generate code following previous thoughts in Table~\ref{tab:prompt_gen_code}.

\begin{table*}[h!]
\centering
\begin{tcolorbox}[
    colback=blue!5!white,
    colframe=black!55!black,
    width=0.98\textwidth,
    title={Prompt for Rethink},
    listing engine=listings,
    boxrule=0.5pt
]
\small
\begin{lstlisting}
Complete the Python program to solve the problem. Remember to contain the complete program including all the imports and function header in your response.

Also some thoughts are included that you can refer to and build upon when writing the code.

Answer with the code ONLY. No other explanation or words attached!

{problem statement}

{thoughts}
\end{lstlisting}
\end{tcolorbox}
\caption{Prompt for generating the code following the thoughts in search methods.}
\label{tab:prompt_gen_code}
\end{table*}

\subsection{Evaluation Prompt}
Besides the normal evaluation on the public test cases, we also develop an LLM-based self-evaluation when the public test cases are all passed. Here we present the prompts in Table~\ref{tab:prompt_eval}.

\begin{table*}[h!]
\centering
\begin{tcolorbox}[
    colback=blue!5!white,
    colframe=black!55!black,
    width=0.98\textwidth,
    title={Prompt for Rethink},
    listing engine=listings,
    boxrule=0.5pt
]
\small
\begin{lstlisting}
{problem statement}

{thoughts}
```python
generated code
```

Above is a Python code problem with the thoughts and code to solve the problem. The code could pass all the example test cases, however, it may or may not be completely correct.

Please evaluate and return the correctness score in range [-1, 1].

Evaluate the correctness of the code and give only ONE evaluation score. 

The code's correctness is whether it can pass all the possible unseen test cases of the problem, not just the given ones.

Example Answers: 
{"evaluation": -0.5, "explanation": The code is far from correct for solving the problem.}
{"evaluation": 0.1, "explanation": The code is not the correct solution but can pass some simple test cases.} 
{"evaluation": 0.85, "explanation": The code can pass most test cases while may fail on some corner cases.} 
{"evaluation": 1.0, "explanation": The generated code is the correct solution that can pass all the possible test cases and strange corner cases too.} 
\end{lstlisting}
\end{tcolorbox}
\caption{Prompt for evaluating the thoughts and code.}
\label{tab:prompt_eval}
\end{table*}

\subsection{Rethink Prompt}
When the generated code following some thoughts doesn't pass some public test cases, we would use the block-level analysis to form the verbal feedback and use it to refine the previous thought, a.k.a, rethink. Here we present the prompt for this operation in Table~\ref{tab:prompt_rethink}.

\begin{table*}[h!]
\centering
\begin{tcolorbox}[
    colback=blue!5!white,
    colframe=black!55!black,
    width=0.98\textwidth,
    title={Prompt for Rethink},
    listing engine=listings,
    boxrule=0.5pt
]
\begin{lstlisting}
{problem statement}
{thoughts}
```python
generated code
```
{verbal feedback}

Based on your previous thoughts and the new experience, please provide a new Thought to replace the previous thought. This new thought should avoid the mistake.

Remember that you only need to provide the thought (one or two sentences) to solve the problem, not the code.

\end{lstlisting}
\end{tcolorbox}
\caption{Prompt for Rethink}
\label{tab:prompt_rethink}
\end{table*}
\subsection{An Example of Accumulated Thoughts}
\label{exm_accumulated_thoughts}
Here we present an example of the thought steps accumulated in one trace of MCTS tree in Table~\ref{tab:example_accumulated_thoughts}.
\begin{table*}[h!]
\centering
\begin{tcolorbox}[
    colback=blue!5!white,
    colframe=black!55!black,
    width=0.98\textwidth,
    title={An Example of Accumulated Thoughts},
    listing engine=listings,
    boxrule=0.5pt
]
\begin{lstlisting}
1-Thought:To maximize the number of edges covered by paths between three distinct vertices in a tree, we can ... yield a higher number of unique edges. 

2-Thought:To enhance the previous thought about leveraging the diameter of the tree, we can implement a breadth-first search (BFS) twice: ... maximizes unique edges covered when chosen as the third vertex. 

3-Thought:To maximize the number of edges covered by paths between three distinct vertices in a tree, we can implement a two-pass BFS ... to maximize edge coverage. 

4-Thought:To further enhance the previous thoughts on maximizing edge coverage among three vertices in a tree, ..., thus ensuring we maximize the unique edges covered by the paths connecting the three vertices. 
\end{lstlisting}
\end{tcolorbox}
\caption{An example of accumulated thought steps. The thoughts are from GPT-4o-mini when solving APPS problem 4000 (1st problem of introductory difficulty):}
\label{tab:example_accumulated_thoughts}
\end{table*}

\subsection{An Example of Verbal Feedback}
\label{exm_vf}
The verbal feedback we constructed contains the detailed block-level analysis of the code. Here we present an example of it.

\begin{table*}[h!]
\centering
\begin{tcolorbox}[
    colback=blue!5!white,
    colframe=black!55!black,
    width=0.98\textwidth,
    title={An Example of Verbal Feedback},
    listing engine=listings,
    boxrule=0.5pt
]
\begin{lstlisting}
(Problem and Code)
import math
def largest_divisor(n: int) -> int:
    """ For a given number n, find the largest number that divides n evenly, smaller than n
    >>> largest_divisor(15)
    5
    """
    for i in range(int(math.sqrt(n)), 0, -1):
        if n % i == 0 and i < n:
            return i
    return 1  # In case n is 1, return 1 as the largest divisor
(Execution Feedback)
The code above fails the given unit test:
assert largest_divisor(15) == 5
# Current Execution Output: 
3. 
(Block-level Analysis)
Here is the code execution trace block by block with the intermediate variable values. Please explain the execution FOR EACH BLOCK and answer whether this block is correct or not. If not, give an explanation on what is wrong. Please wrap your response into a JSON object that contains keys `block` with the name of each block, key `correct` with value False or True, and key `explanation` with an explanation on the bug. 
Example Answers:
{"block": "BLOCK-1", "correct": "True", "explanation": "The block initializes variable `a` and `b`."}
{"block": "BLOCK-2", "correct": "False", "explanation": "The block is incorrect because the code does not add the two integers together, but instead subtracts the second integer from the first. To fix this issue, we should change the operator from `-` to `+` in the return statement. This will ensure that the function returns the correct output for the given input."}
[BLOCK-0]
    # n=15
    for i in range(int(math.sqrt(n)), 0, -1):
    # n=15	i=3
[BLOCK-1]
        # n=15	i=3
        if n % i == 0 and i < n:
        # n=15	i=3
[BLOCK-2]
            # n=15	i=3
            return i
            # n=15	i=3	_ret=3
            
# LLM-analysis
```json
[
    {"block": "BLOCK-0", "correct": "True", "explanation": "The block initializes the loop with the correct starting point, which is the integer part of the square root of n."},
    {"block": "BLOCK-1", "correct": "False", "explanation": "The block checks if 3 is a divisor of 15, which is correct, but we are not immediately returning the largest divisor. The loop continues, and the next iteration will check smaller values of i."},
    {"block": "BLOCK-2", "correct": "False", "explanation": "The block incorrectly returns 3 instead of the larger valid divisor. Since the loop searches from the square root downwards, it should continue checking until it finds the next valid divisor, which is 5. The return statement needs to occur outside of this loop and only when the largest divisor is found."}
]
```
\end{lstlisting}
\end{tcolorbox}
\caption{An example of verbal feedback.}
\label{tab:example_vb}
\end{table*}

\section{Algorithm}
\label{ag:algs} %

We present the detailed procedure of \our{} in pseudocode in Algorithm~\ref{alg:ouralg}.

\setlength{\intextsep}{0pt} %
\begin{figure}
\begin{minipage}{\textwidth}
\begin{algorithm}[H]
\caption{The \our{} algorithm.}
\label{alg:ouralg}
\small{
\begin{algorithmic}[1]
\Require $root$: the problem description;
$c$: P-UCB exploration parameter;
$k$: the maximum number of children of any node;
$a, b$: the reward weights of the pass rate and the LLM evaluation.
\State $program\_dict = \texttt{\MakeUppercase{dictionary}}()$
\State $verbal\ feedback\ f = \texttt{\MakeUppercase{empty}}$
\For{$i \gets 1, 2, \dots, max\_rollouts$}
    \State $node \gets root$
    \State {\color{ForestGreen} {\# Selection}}
    \While{$|node.children| > 0$}
        \State $node \gets {\texttt{\MakeUppercase{p\_ucb\_select}}}(node.children, c)$
    \EndWhile \label{line:end_of_selection}
    \State {\color{ForestGreen} {\# Expansion}}
    \State $next\_thoughts \gets \texttt{\MakeUppercase{top\_k}}(node, k)$

    \For{$next\_thought \in next\_thoughts$}
        \State $next\_state \gets \texttt{\MakeUppercase{concat}}(node, next\_thought)$
        \State Create a node $new\_node$ for $next\_state$
        \State Add $new\_node$ to the children of $node$
    \EndFor \label{line:add_new_node}
    \State {\color{ForestGreen} {\# Evaluation}}
    \State $C \gets \texttt{\MakeUppercase{generate}}(node)$
    \label{line:generate_program_list}
    \State $v^{\text{test}}, f \gets \texttt{\MakeUppercase{get\_pass\_rate}}(p)$ 
    \State $v^{\text{llm}}, f \gets \texttt{\MakeUppercase{get\_llm\_eval}}(p)$ 
    \State $program\_dict[C] = r = a*v^{\text{test}}+b*v^{\text{llm}}$
    \If{$v^{\text{test}} = 1$}
        \State $program\_dict[C] = r = a*v^{\text{test}}+b*v^{\text{llm}}$
    \Else
        \State $program\_dict[C] = r = v^{\text{test}}$
    \EndIf
    \State {\color{ForestGreen} {\# Backpropagation}}
    \State Update and the values of $node$ and its ancestors in the tree with $r$
    \State {\color{ForestGreen} {\# Rethink}}
    \If{$v^{\text{test}} \neq 1$}
        \State $node.thought = \texttt{\MakeUppercase{rethink}}(node, f)$
        \State $next\_thoughts = \texttt{\MakeUppercase{rethink\_next}}(node, k, f)$
        \State $C = \texttt{\MakeUppercase{re-generate}}(node)$
        \State $r = \texttt{\MakeUppercase{re-evaluation}}(C)$
        \State $program\_dict[C] = r$
    \EndIf
\EndFor
\State \textbf{return} program in $program\_dict$ with the highest reward
\end{algorithmic}
} %
\end{algorithm}
\end{minipage}
\end{figure}

\end{document}